\documentclass[a4paper,11pt]{article}
\pdfoutput=1 
\usepackage{jinstpub} 

\usepackage{subfig}

\title{\boldmath Characterization of new eco friendly gas mixtures based on HFO for RPCs }

\author[a,b,1]{G. Proto,\note{Corresponding author.}}
\author[a,b]{G. Aielli,}
\author[a,b]{E. Alunno Camelia,}
\author[a]{P. Camarri,}
\author[b]{R. Cardarelli,}
\author[a,b]{A. Di Ciaccio,}
\author[a]{L. Di Stante,}
\author[b]{B. Liberti,}
\author[c]{A. Paoloni,}
\author[b]{E. Pastori,}
\author[b]{L. Pizzimento,}
\author[b]{A. Rocchi,}
\author[a,b]{R. Santonico,}
\author[a]{E. Tusi}

\affiliation[a]{University of Rome Tor Vergata,\\Via della Ricerca Scientifica 1, Rome, Italy}
\affiliation[b]{INFN Rome Tor Vergata,\\Via della Ricerca Scientifica 1, Rome, Italy}
\affiliation[c]{INFN Laboratori Nazionali di Frascati (LNF),\\Via Enrico Fermi, 54, Italy}

\emailAdd{giorgia.proto@roma2.infn.it}

\abstract{The ATLAS RPC standard mixture, mainly based on C$_{2}$H$_{2}$F$_{4}$, has a high Global Warming Potential (GWP) and therefore the search for RPC eco friendly gases is mandatory. In this work we present the results on the detector performances in terms of efficiency, prompt and ionic charge, with different gas mixtures.}


\proceeding{15$^{\text{th}}$ Workshop on Resistive Plate Chambers and related detectors\\
  10-14 February 2020 \\
  Rome}

\begin{document}
\maketitle
\flushbottom

\section{Introduction}
\label{sec:intro}
The RPC standard mixture, composed by C$_{2}$H$_{2}$F$_{4}$/i-C$_{4}$H$_{10}$/SF$_{6}$/=94.5/5/0.5, is presently used in several experiments with a high radiation environment. Unfortunately it has a high Global Warming Potential of about 1490, mainly due to the C$_{2}$H$_{2}$F$_{4}$ ( SF$_{6}$ gives a minor contribution of $\sim$ 90). The use of this mixture could be problematic in the near future if additional restrictions on the greenhouse gas emissions will be imposed.\\ The main problem in searching for alternative mixtures is to maintain good avalanche saturation properties and the separation between avalanche operation mode and streamer contamination over a large electric field range ($\sim$ 1 kV).\\ For what concerns the new generation of RPCs that are planned for the LHC upgrade, the new detector layout together with the development of new Front-End electronics give the possibility to work with new mixtures with lower GWP  ~\cite{g,h,i,m,n}, based on HFO, CO$_{2}$, i-C$_{4}$H$_{10}$ and SF$_{6}$. Instead the RPCs already installed in the experiment that can't be upgraded, can not work with these new mixtures.\\ 
With HFO-based new gas mixtures, the avalanches are wider and carry more charge and the transition to streamers passes through multi-avalanches events with a high charge content. The charge distributions are broader and the saturated-avalanche regime is achieved only when the efficiency is still too low.\\ The goal of this study is to identify the best new mixture and to optimize the  percentages of the various component inside the mixture.
 \section{Experimental setup}
 \label{sec:setup}
\begin{figure}[h]
 	\centering
 	\includegraphics[width=.4\textwidth]{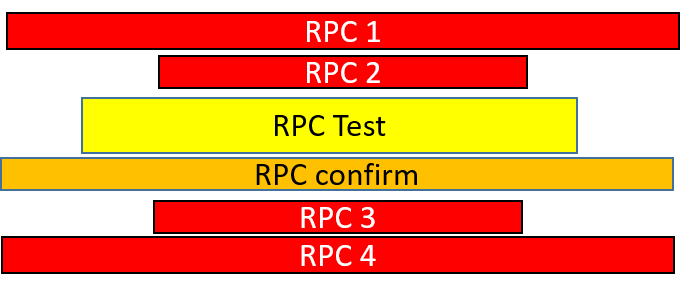}
 	\caption{\label{fig:setupexp} Schematic view of the experimental setup.}
 \end{figure}

 A schematic view of the experimental setup is shown in Figure \ref{fig:setupexp}.
 
An RPC with 2 mm gas gap and two 1.8 mm thick electrodes (ATLAS-like) has been used in this test. The detector dimensions are 57x10 cm$^{2}$. The prompt induced signal is read out on both sides of a single strip line with the oscilloscope without signal amplification. One of the two signals is acquired with the maximum sensibility of the oscilloscope to optimize  the analysis for the avalanche, while the other is acquired with a variable scale to study the streamers. The oscilloscope has a bandwidth of 3 GHz and a sampling velocity of 20 ${Gs}/{s}$ and the time window acquired for the prompt signal is 200 ns.\\  
The ionic signal is read out on a resistance of about 10 k$\Omega$ on the ground graphite electrode over a time window of 100 $\mu$s.\\ The trigger system consists in four RPCs (see Figure \ref{fig:setupexp}) and a 1 mm gas gap RPC has been used as time reference and confirmation for the efficiency measurement.
 \section{Analysis Parameters Definition}
A signal is considered efficient if it crosses an amplitude threshold, defined as the 5*RMS of the background amplitude distribution which is measured in a time window of 20 ns just before the trigger signal. No Front-End electronics has been used, and the lowest equivalent threshold is $\sim$ 2.2 mV( highest oscilloscope sensitivity), which is substantially higher than the one of the ATLAS RPCs ($\sim$ 1.2 mV).\\The avalanche charge, the total prompt charge and the ionic charge are measured for each acquired waveform. The avalanche charge is defined as the integrated charge in 10 ns around the first peak. The total prompt charge is the  integrated charge in the whole time window of 180 ns (the first 20 ns are used for the background measurement), so it includes also the multi-avalanches and the streamers contributions. The ionic charge is measured in the full time window of 80 $\mu$s after trigger. The window of 20 $\mu$s before the trigger has been used for the background measurement. \\
The delivered charge in HFO-based mixtures takes an important contribution from multi-avalanches, therefore, instead of streamer, we prefer to define an \emph{extra-charge} , due to all signals observed in 180 ns. A threshold of 20 pC has been fixed to discriminate what we call an \emph{extra-charge} event. This parameter includes the contributions of multi-avalanches signals, very wide avalanches and streamers.    
\section{Measurements Strategy and Experimental Results}
\begin{figure}[h]
	\includegraphics[width=1.2
\textwidth]{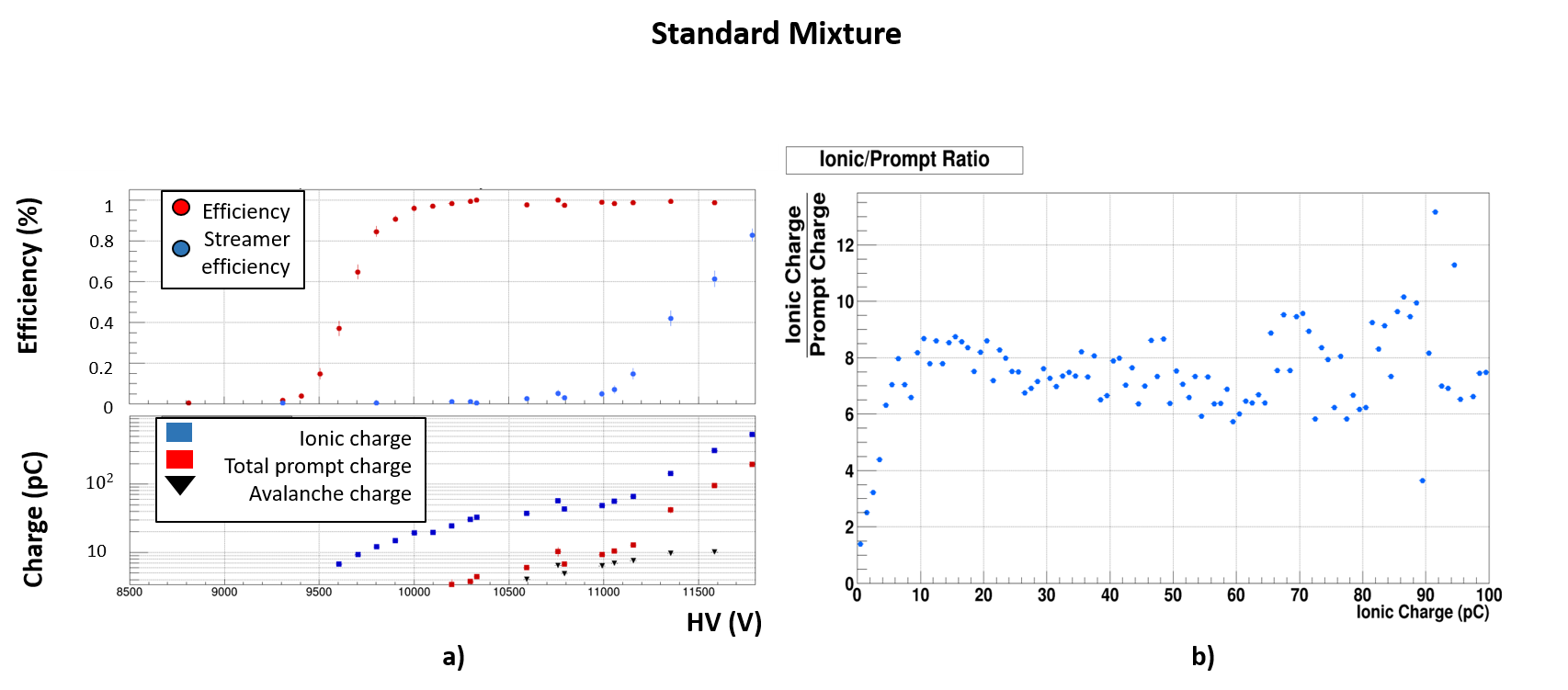}
	\caption{\label{fig:mixstandard} Standard Mixture results.\textbf{ a-top)} Efficiency (red) and Streamer probability (blue) as a function of the high voltage.\textbf{ a-bottom)} Ionic charge (blue), total prompt charge (red) and avalanche charge (blue) as a function of the high voltage.\textbf{ b)} Ionic/prompt ratio as a function of the ionic charge }
\end{figure}

In order to study the properties of the different gas components, a series of measurements have been performed starting from the mixture composed by C$_{3}$H$_{2}$F$_{4}$(HFO-1234ze)/CO$_{2}$/i-C$_{4}$H$_{10}$/SF$_{6}$ =38/ 56/5/1. In the first series the CO$_{2}$/SF$_{6}$ ratio has been kept constant, while HFO/i-C$_{4}$H$_{10}$ percentages have been changed. In the second series the i-C$_{4}$H$_{10}$/SF$_{6}$ fraction is fixed and the percentages of the main components of the mixture vary.\\Finally, a different new type of HFO, C$_{3}$H$_{2}$ClF$_{3}$ (HFO-1233zd), has been tested. The performance of each gas mixture has been studied in terms of efficiency, avalanche-extra charge separation and total charge delivered in the detector together with the ionic/prompt charge ratio. For all measurements the standard mixture has been kept as a reference. In Figure \ref{fig:mixstandard} the results for the standard mixture are summarized.\\ The efficiency curve shows that the voltage gap between the efficiency plateau and the streamer appearance  is of $\sim$ 1 kV (Figure \ref{fig:mixstandard}.a top). The ionic to prompt charge ratio as a function of the ionic charge reaches at 10 kV (efficiency plateau) the constant value of  8, for an average ionic charge value of $\sim$ 11 pC (Figure \ref{fig:mixstandard}.b). All the high voltage points are considered in the measurement of the ionic and prompt charge.  


\subsection{Mixtures with HFO1234ze/CO$_{2}$ /i-C$_{4}$H$_{10}$/SF$_{6}$}
The first measurment has been performed with  the mixture composed by HFO1234ze/CO$_{2}$/i-C$_{4}$H$_{10}$/SF$_{6}$ = 38/56/5/1, with the CO$_{2}$ as the main component.\\ In general we would like to work with as much CO$_{2}$ as possible, because the HFO plays the role of \emph{quencher} and it moves the working point towards very high voltages. A greater fraction of HFO is expected to show lower extra charge probability and lower current flowing inside the detector.\\Two sets of measurements have been performed, one to study the quenching components and the other to study the ratio between the main components of the mixture.
\subsubsection{Performance with HFO1234ze/ i-C$_{4}$H$_{10}$ variable ratio}
\begin{figure}[h]
	\centering 
	\includegraphics[width=1.\textwidth]{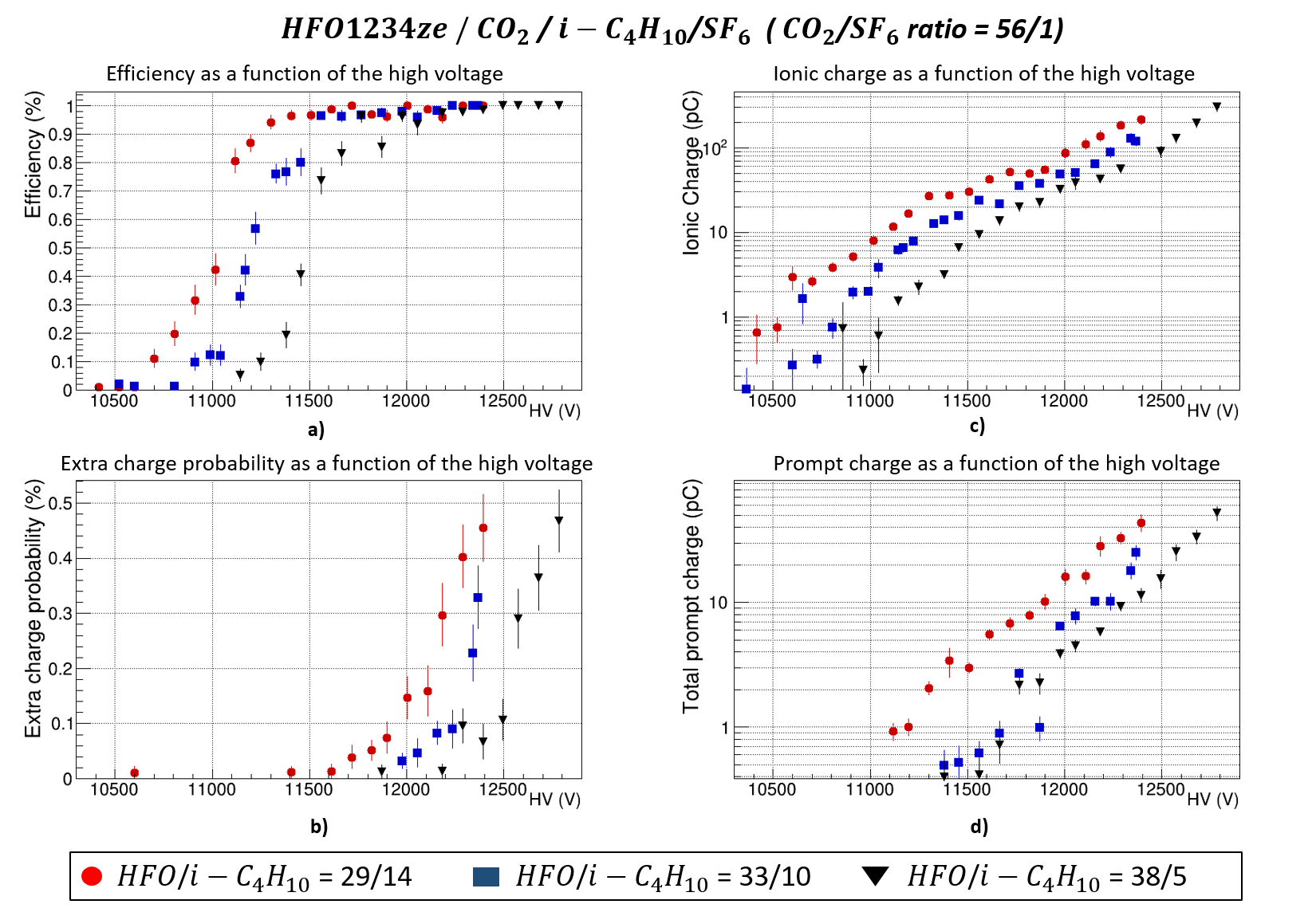}
	\caption{\label{fig:firstset}Efficiency (a), extra charge probability (b), ionic charge (c) and total prompt charge (d) as a function of the high voltage at fixed 56/1 CO$_{2}$/SF$_{6}$ ratio for the three mixtures: HFO/i-C$_{4}$H$_{10}$= 29/14 (red points), 33/10 (blue squares) and 38/5 (black triangles).}
\end{figure}

In this section the CO$_{2}$ and SF$_{6}$ concentrations are fixed at the ratio 56/1, while the HFO percentage is lowered in favour of i-C$_{4}$H$_{10}$. Three mixtures are studied: HFO/i-C$_{4}$H$_{10}$ = 38/5, 33/10, 29/14.\\The results are reported in Figure \ref{fig:firstset}.\\
We can notice that the addition of 10\% of i-C$_{4}$H$_{10}$ lowers the operating voltage of about 600 V and the avalanche-extra charge separation is $\sim$ 600 V for all the three mixtures (Figure \ref{fig:firstset}.a-\ref{fig:firstset}.b). If we look at the total prompt charge we can see that the average value is the same ($\sim$ 4 pC) for the three mixtures at the respective working points (Figure \ref{fig:firstset}.d). The same happens for the ionic charge, which represents the total charge delivered inside the detector and takes the value of $\sim$ 30 pC at the respective working points (Figure \ref{fig:firstset}.c).\\ In conclusion there is no effect in increasing i-C$_{4}$H$_{10}$ in terms of avalanche-extra charge separation and total charge delivered at the working point, but only in terms of operating voltage which is lower in the mixture with less HFO.
\subsubsection{Performance with HFO1234ze/CO$_{2}$ variable ratio}
\begin{figure}[h]
	\centering 
	\includegraphics[width=1.\textwidth]{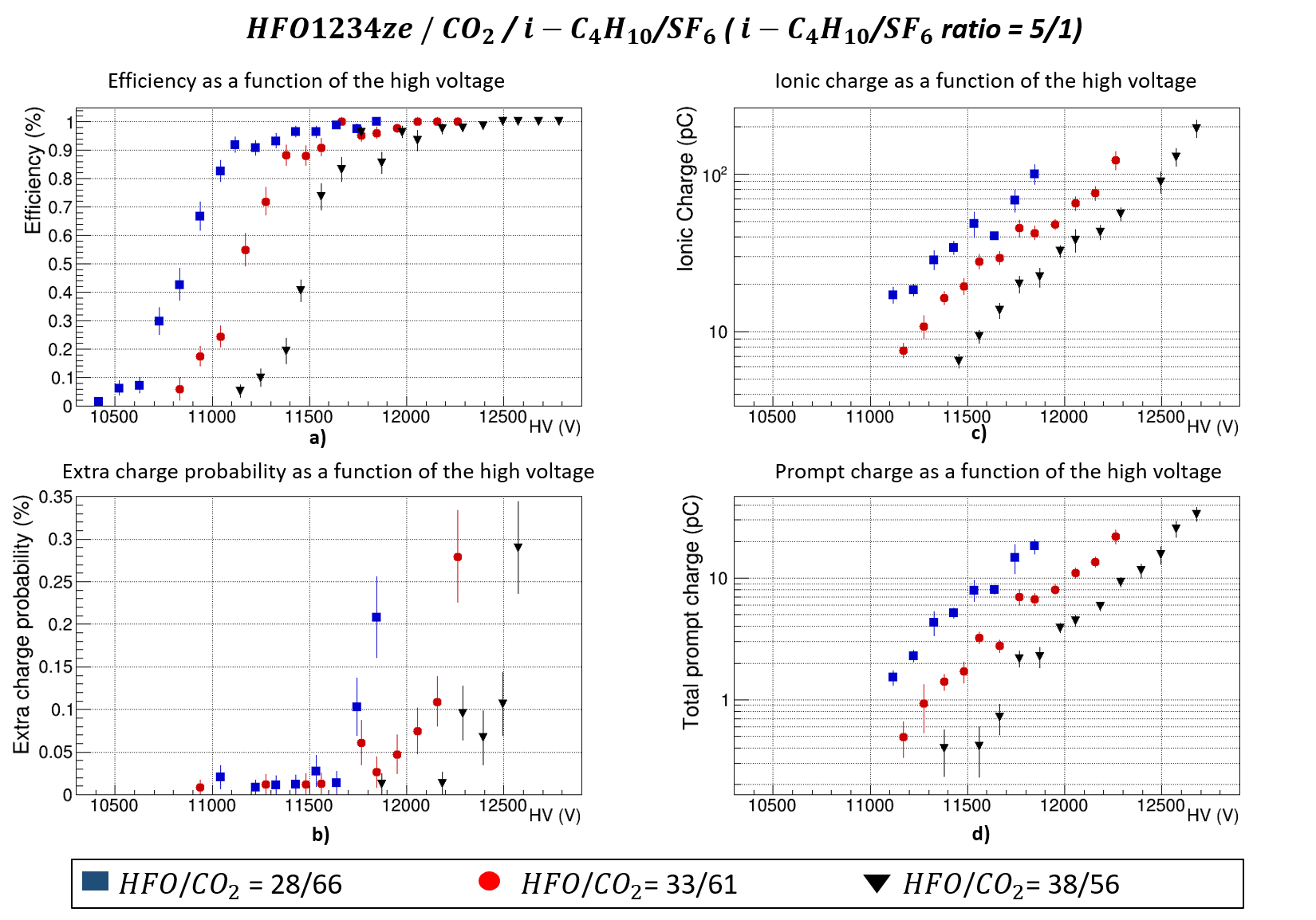}
	\caption{\label{fig:secondtest} Efficiency (a), extra charge probability (b), ionic charge (c) and total prompt charge (d) as a function of the high voltage at fixed 5/1 i-C$_{4}$H$_{10}$/SF$_{6}$ ratio for the three mixtures: HFO/CO$_{2}$= 28/66 (blue squares), 33/61 (red points) and 38/56 (black triangles).}
\end{figure}

In this section the fraction i-C$_{4}$H$_{10}$/SF$_{6}$ has been kept constant to the ratio 5/1 and the concentration of CO$_{2}$ has been increased in steps of 5\%. The reported HFO/CO$_{2}$ ratios are 38/56, 33/61 and 28/66.\\ In Figure \ref{fig:secondtest} the results are reported. It can be seen that the reduction of HFO moves the working point to lower voltages of about 400 V every 5\% of HFO reduction (Figure \ref{fig:secondtest}.a). The total prompt charges for the three gas mixtures at the working point are consistent with each other ($\sim$ 3 pC) but the total ionic charge is higher in the mixture with more HFO ($\sim$ 30 pC against $\sim$ 20 pC for the other two) as you can see in Figures \ref{fig:secondtest}.c-\ref{fig:secondtest}.d.\\ For what concerns the avalanche-extra charge separation, the range is larger in the mixture with more HFO (Figure \ref{fig:secondtest}.b).
\subsection{Mixtures with HFO1233zd/CO$_{2}$ /i-C$_{4}$H$_{10}$/SF$_{6}$}
In this section the results related to a different type of HFO, the HFO1233zd~\cite{l}, are reported. This molecule, C$_{3}$H$_{2}$ClF$_{3}$, has an extra Chlorine respect to the previous one and one Fluorine less.
Similar molecules containing two different halogen atoms, were already tested as RPC gas components. CF$_{3}$Br was shown to have a relevant effect in reducing the streamer size in RPC operating in streamer mode~\cite{z}. Moreover, the pure avalanche mode without any streamer contamination was first observed in RPCs filled with pure CF$_{3}$Br.  More recently, the CF$_{3}$I molecule was successfully tested as an alternative RPC gas~\cite{n}. Unfortunately, none of these molecules is suitable for very large size RPC systems, where industrial gases are needed. Indeed CF$_{3}$Br has a very high ozone depletion potential and CF$_{3}$I is very poisonous.\\ C$_{3}$H$_{2}$ClF$_{3}$ is an industrial gas and does not have the above problems and the test made with the other two-halogen molecules suggests that it could be suitable to suppress the streamer appearance and to keep the avalanche size to modest values.
Moreover the quenching power of this gas, HFO 1233zd, is expected to be stronger than that of HFO-1234ze. The properties of this gas suggest that it might replace the SF$_{6}$.\\ A test with a new gas mixture composed by HFO-1233zd/CO$_{2}$ /i-C$_{4}$H$_{10}$ without the presence of SF$_{6}$ has been performed, in order to study if this gas could give the same benefits of the SF$_{6}$.   
\subsubsection{ Performance with mixtures with and without SF$_{6}$}
\begin{figure}[h]
	\centering 
	\includegraphics[width=1.\textwidth]{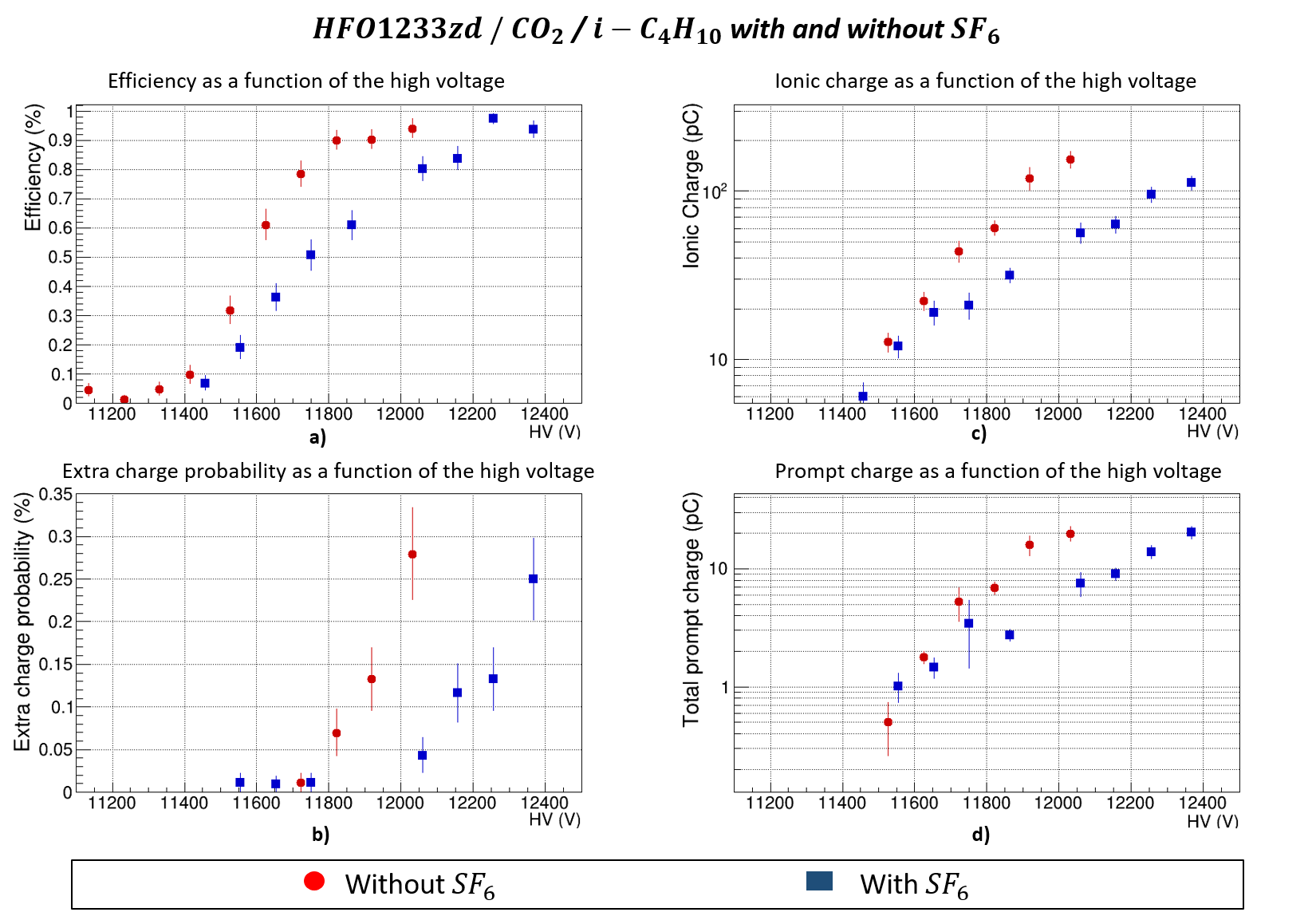}
	\caption{\label{fig:hfonew1}Efficiency (a), extra charge probability (b), ionic charge (c) and total prompt charge (d) as a function of the high voltage for the two mixtures: HFO1233zd/CO$_{2}$/i-C$_{4}$H$_{10}$ without SF$_{6}$ (red points) and with SF$_{6}$ (blue squares).}
\end{figure}

\begin{figure}[!hb]
	\centering 
	\includegraphics[width=.5\textwidth]{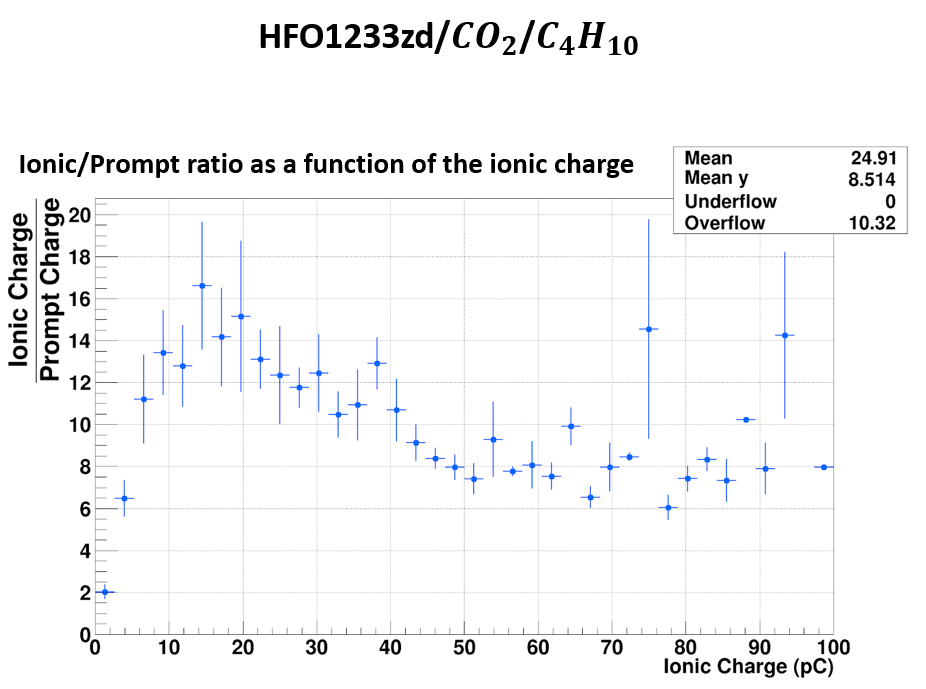}
	\caption{\label{fig:ionicprompt} Ionic/Prompt ratio as a function of the ionic charge for the HFO1233zd/CO$_{2}$ /i-Butane Mixture .}
\end{figure}

The mixtures under study are composed by HFO-1233zd/CO$_{2}$ /i-C$_{4}$H$_{10}$=15/80/5 with and without the addition of 1\% of SF$_{6}$.
The experimental results are reported in Figure \ref{fig:hfonew1}.\\It can be seen that the mixture without SF$_{6}$ has an operating voltage $\sim$ 300-400 V (at 90\% efficiency) lower and shows a steeper rise. This is an expected behaviour~\cite{c} and can be explained with the very high electronegativity of this molecule. In fact the simultaneous presence of the HFO and SF$_{6}$ breaks the avalanche growth and makes harder the rise of the efficiency curve.\\The total prompt charge at the plateau knee (11800 V and 12100 V) is higher in the mixture with the SF$_{6}$ ($\sim$ 6 pC against 8 pC) and the same is true for the ionic charge delivered in the detector ($\sim$ 40 pC against 60 pC). Therefore, when SF$_{6}$ is added, it finds a very electronegative environment and the resulting effect is the premature break of the avalanche growth and not the streamers suppression.
In Figure \ref{fig:ionicprompt} the ionic/prompt charge ratio as a function of the ionic charge is shown.  

The ratio is $\sim$ 8 at the mean value of the ionic charge at the operating voltage as in the standard mixture, but this plateau value is reached only when the extra charge events occur. In addition, even if the ratio has the same value for both mixtures, the mean value of the two charges is twice in the HFO-based mixture.

\begin{figure}[!ht]
	\centering 
	\includegraphics[width=1.2\textwidth]{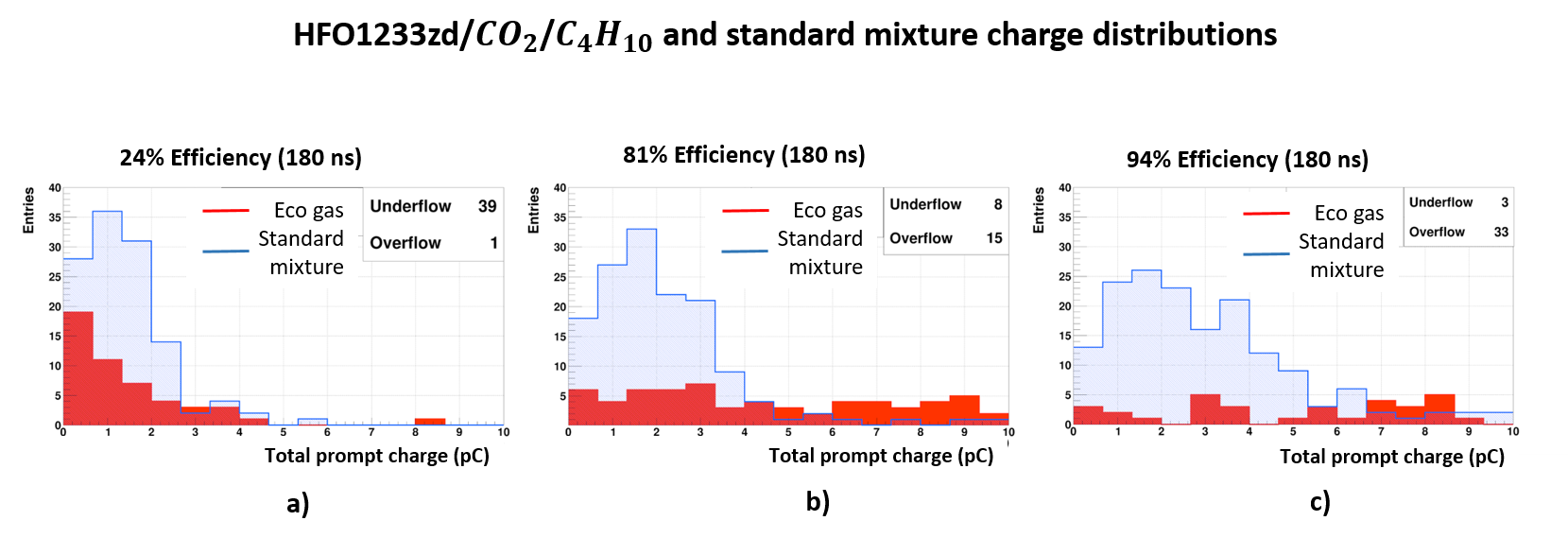}
	\caption{\label{fig:dis} Total prompt charge distribution at  24\% (a), 81\% (b) and 94\% (c) of efficiency for the standard mixture (blue) and HFO1233zd/CO$_{2}$ /i-Butane Mixture (red).}
\end{figure}

Figure \ref{fig:dis} shows the comparison between the total prompt charge distributions of the standard mixture and the HFO1233zd/CO$_{2}$ /i-C$_{4}$H$_{10}$ mixture at different values of the efficiency.\\
  At lower efficiency the two distributions are concentrated near the ideal line which separate the signal charge and the background charge. When efficiency rises the total prompt charge distribution for the standard mixture shows the typical trend of saturated avalanche regime, on the opposite the HFO mixture has a broader distribution, causing an higher average working prompt charge value (see Figure \ref{fig:dis}).\\\\\\

\section{Conclusions}
In this work we made a systematic performance study of a small size RPC filled with four gas mixture components C$_{3}$H$_{2}$F$_{4}$(HFO1234ze)/CO$_{2}$/i-C$_{4}$H$_{10}$/SF$_{6}$, using the following different proportions among them: 29/56/14/1, 33/56/10/1, 38/56/5/1, 28/66/5/1 and 33/61/5/1. We obtained the following results:
\begin{itemize}
	\item {Decreasing i-C$_{4}$H$_{10}$ in the HFO/i-Butane ratio produces a decrease of the operating voltage of about $\sim$ 600 V but does not change the delivered charge, which is the same in the three mixtures and has a value of $\sim$ 30 pC. Also the avalanche- extra charge separation is the same in the three mixtures;}
	\item{Decreasing the HFO in the HFO/ CO$_{2}$ ratio has the effect to decrease both the operating voltage and the delivered charge. The voltage decrease is $\sim$ 800 V and the delivered charge decrease is $\sim$ 10 pC. This suggests that HFO concentration should be lowered below the 28\% value tested in this study. The avalanche-extra charge separation range is larger in the mixture with more HFO;}
\end{itemize}
We tested also another HFO gas, C$_{3}$H$_{2}$ClF$_{3}$ (HFO1233zd), that has never tested before. This new component has much stronger quenching power than the previous HFO and this suggests that it might replace the SF$_{6}$. However this possibility requires more studies.

\end{document}